# Emergence of the electronic states driven by quantum charge fluctuations in electron-doped cuprate superconductors

Hiroshi Yamaguchi[1], Yudai Miyai[1], Yuki Tsubota[1], Yogendra Kumar[2], Masashi Arita[2], Hitoshi Sato[1,2], Dongjoon Song[3], Kiyohisa Tanaka[4], Kenya Shimada[1,2,5,6] & Shin-ichiro Ideta[1,2]*

[1]*Graduate School of Advanced Science and Engineering, Hiroshima Univ., Higashi-Hiroshima 739-0046, Japan*

[2]*Research Institute for Synchrotron Radiation Science (HiSOR), Hiroshima Univ., Higashi-Hiroshima 739-0046, Japan*

[3] *Quantum Matter Institute, University of British Columbia, British Columbia, Vancouver, V6T 1Z4, Canada*

[4] *UVSOR-III Synchrotron, Institute for Molecular Science., Okazaki, Aichi 444-8585, Japan*

[5]*Research Institute for Semiconductor Engineering, (RISE), Hiroshima Univ., Higashi-Hiroshima 739-8527, Japan*

[6] *International Institute for Sustainability with Knotted Chiral Meta Matter (WPI-SKCM²), Higashi-Hiroshima 739-8531, Japan*

* email: idetas@hiroshima-u.ac.jp

**Abstract**

The origin of electron−boson interactions is central to understanding high-$T_c$ superconductivity in cuprates. While phonons and magnetic fluctuations are widely considered as candidates for mediating electron pairing, the role of charge fluctuations, one of the fundamental electronic degrees of freedom, remains unclear. Here, we investigate the electronic structure of the electron-doped cuprate using angle-resolved photoemission spectroscopy and angle-resolved and resonant inverse photoemission spectroscopies, which reveal the electronic structure in the occupied and unoccupied states, respectively. From these spectroscopic experiments, we find emergent spectral features on both occupied and unoccupied states that are consistent with excitations driven by quantum charge fluctuations. The results obtained in this study offer direct experimental insight into charge fluctuations in cuprates, thereby paving the way towards clarifying the electronic structure of cuprates to understand the mechanism of high-$T_c$ superconductivity.

**Introduction**

In the parent compounds of copper oxide high-$T_c$ superconductors for both electron- and hole-doped systems, the antiferromagnetic order is gradually suppressed by carrier doping, leading to the emergence of high-$T_c$ superconductivity [1, 2]. The mechanism of superconductivity in the $CuO_2$ planes arises from the interactions between Bogoliubov quasiparticles (QPs) and various bosonic excitations including spin fluctuations, phonons, or instantaneous spin exchange processes. Although understanding the origin of these bosons has been a central issue in unraveling the mechanism of high-$T_c$ superconductivity in cuprates, the origin of the relevant bosonic modes remains elusive despite extensive experimental and theoretical efforts [3-14].

It is well known that materials exhibit intriguing physical properties owing to the complex interplay of various internal degrees of freedom, such as superconductivity, metal−insulator transitions, and ferroelectricity.

Recently, the mechanism underlying high-$T_c$ superconductivity in cuprates has drawn significant interest, particularly regarding its correlation with charge degrees of freedom. X-ray scattering and scanning tunneling microscopy studies have reported the observation of charge density waves (CDWs) that locally break the translational symmetry in hole-doped cuprates in the underdoped regime [15–17]. Additionally, anisotropy in transport properties suggests the presence of electron nematicity, which breaks the $C_4$ rotational symmetry [18,19]. These statically ordered phases, along with antiferromagnetism and superconductivity, are crucial for understanding the microscopic electronic structure of cuprates [20]. In contrast, the significance of dynamic charge fluctuations has recently been revealed in details by resonant inelastic X-ray scattering (RIXS) experiments [21–28], which directly observe charge excitations in the $CuO_2$ planes. These findings provide clear evidence for the dual nature of the electronic structure in cuprate superconductors: they form localized orders such as static CDWs, but might also exhibit acoustic plasmon excitations arising from long-range Coulomb interactions. In addition, the interaction between low-energy plasmons and the electronic structure of cuprates is expected to create QP dispersions in the occupied and unoccupied states [29-33].

To understand the effect of charge fluctuations on the electronic structure of cuprates, Yamase *et al.* recently employed the extended *t-J-V* model and showed that it could quantitatively reproduce the dispersion relations and energy gaps of plasmons observed in RIXS experiments [21, 22, 31]. By including on-site charge and bond-charge fluctuations, they predicted that the interaction between charge fluctuations and the electronic states on the $CuO_2$ planes leads to the emergence of QP dispersions in both occupied and unoccupied states [30–33]. The formation of the QP dispersion generated by coupling to charge fluctuations has been observed in weakly correlated systems, such as graphene [34, 35], two-dimensional GaAs/$Al_xGa_{1-x}As$ heterostructure [36], and $SrIrO_3$ films [37]. The electronic structure derived from charge fluctuations in strongly correlated systems such as cuprate superconductors remains unclear. Since previous theories are reported that the intensity of these electronic structures induced by charge fluctuation is weak [30-33], the observations of the emergent QP bands by angle-resolved photoemission spectroscopy (ARPES) would be intrinsically challenging. However, understanding the interplay between charge fluctuations and electronic states in high-$T_c$ cuprate superconductors is expected to provide valuable insights into the mechanism underlying high-$T_c$ superconductivity.

To investigate how charge fluctuations in cuprates contribute to the electronic structure, we performed ARPES and angle-resolved inverse photoemission spectroscopy (AR-IPES) to probe the occupied and unoccupied states of the electron-doped high-$T_c$ cuprate superconductor $Nd_{2-x}Ce_xCuO_4$ ($x$ ~ 0.15, NCCO), respectively. Additionally, we performed density functional theory (DFT) and dynamical mean-field theory (DMFT) calculations for 49-atom $Nd_{13}Ce_1Cu_7O_{28}$ supercell, corresponding to $x = 1/7 \sim 0.143$. This composition closely approximates that of the sample. The calculation provides us to assess the contributions of localized Nd/Ce 4$f$ states to the unoccupied electronic structure and to help distinguish them from the emergent QP bands observed experimentally. The calculations are therefore used here primarily as a reference for rare-earth-derived states, rather than as a quantitative description of the low-energy charge-fluctuation electronic structure in NCCO (see Methods for details). In this study, the electronic structure was observed in the ARPES data, appearing from approximately the Fermi level ($E_F$) to –1 eV over a wide momentum range. In addition, the electronic structure of the unoccupied states was successfully observed by AR-IPES and resonant IPES (RIPES) on NCCO, and unexpected spectral features were observed. A comparison of the present experimental results with those of previous studies [30–33] and the present DFT+DMFT calculations suggests that the observed spectral features are likely induced by charge fluctuations. As this study provides the first experimental evidence of the coupling between quantum charge fluctuations and the electronic structure in cuprates, revealing emergent QP band structures, it offers important constraints for future theoretical model.

## Results

Figure 1 shows the ARPES intensity plots of NCCO along the nodal direction (0, 0) – (π, π). In Fig. 1(a), a weak but clearly resolved nodal band crosses $E_F$, which primarily originates from the Cu $3d_{x^2-y^2}$ and O $2p_{x/y}$ orbitals [38]. It is important to note that the ARPES spectral intensity of NCCO near $E_F$ is strongly dependent on the photoemission matrix element, such that clear band structures are observed only when using photon energies around 16 – 17 eV [39]. Figure 1(b) shows a strong spectral intensity at deeper energies from the range of –2 to –7 eV owing to the Cu 3*d* and O 2*p* bands [38]. In addition to this strong intensity, we found that a hump structure is observed below the peak of the Cu 3*d* as shown by energy-distribution curves (EDCs) in Figs. 1(b). To investigate the momentum dependence of the electronic structure, we showed EDCs in Figs. 2(a1) – 2(a3) and performed a second–derivative analysis of the energy–momentum (*E-k*) plots along the energy direction. The analyzed *E-k* plots for the momentum directions (0, 0) – (π, π), (0, 0) – (π, 0), and (π, 0) – (π, π) are shown in Figs. 2(b1) – 2(b3) with the DFT+DMFT calculation in the occupied state. The intensity of the nodal QP band structure below $E_F$ in Figs. 2(a1) and 2(b1) has been observed, and the parabolic QP band in Figs. 2(a3) and 2(b3) represents the antinodal band structure. The flat bands below $E_F$ in Figs. 2(a2) and 2(b2) are also observed probably due to the Cu 3*d* and O 2*p* orbitals in the presence of the self-energy [31]. In the present DFT+DMFT calculation, the effect of the band renormalization is not taken into account. In addition to the main electronic structure near $E_F$, we also observed that the electronic structure is present from –1 eV to the $E_F$ ($\omega_1$). Although the hump structure is marked by a purple-circled marker, its intensity is weaker than that of the other electronic structure marked by the cross markers and dotted curves originating from the Cu 3*d* and O 2*p* orbitals. The electronic structures below -2 eV in the experiment are well reproduced by the present calculations. According to previous large-*N* theories reported by Yamase *et al.*, emergent bands originating from optical plasmons, incoherent plasmarons, are expected to appear at –1 eV below $E_F$ in electron-doped cuprates, and these dispersive features resemble the main band crossing $E_F$ [31, 33]. In the present ARPES results, the energy position of the hump structure shown in Fig. 2 agrees with the energy of the theoretical prediction. Here, the QP feature observed by ARPES around $\omega_1$ appears only as faint shadows, but the spectral intensity predicted by the theory also shows a weak intensity compared with the main QP bands [31]. Therefore, the broad spectral weight extending over a wide range of momentum can be interpreted as the result of incoherent plasmarons. Although the intensity of the QP bands in the energy range of $\omega_1$ is small, such weak spectral weight is typical of emergent bands coupled with charge fluctuations [40].

We also investigate whether the high-energy kink (waterfall) could be the origin of the hump structure. The high-energy kink is a well-known anomaly characterized by a strong suppression of spectral intensity around –0.5 eV, followed by a recovery near –1 eV [39, 41, 42]. This high-energy anomaly has been reported in both hole- and electron-doped cuprates along the nodal direction and is considered a universal feature of cuprate superconductors. In the present study, since the hump structure appears over a wide momentum range as shown in Fig. 2, we therefore conclude that the hump structure at $\omega_1$ is probably not caused by the signature of the high-energy kink.

In addition to the ARPES experiments, an AR-IPES experiment was performed on NCCO. While ARPES and theoretical approaches have significantly advanced the understanding of the occupied electronic structures in electron-doped cuprates [43, 44], the many-body relaxation processes associated with electron addition remain a comparatively less explored. IPES is a powerful experimental tool for investigating the electronic structure in the unoccupied state and a complementary technique for ARPES. As illustrated in Fig. 3, the AR-IPES spectra have been measured along the same momentum directions as those employed in the ARPES experiments. As shown in Fig. 3(a), the electronic structure has been observed at ~2–4 eV ($\omega_2$, highlighted in red) for all momentum directions. Moreover, by applying second-derivative analysis to the raw spectra, two or three additional structural features have been identified above 5 eV in Fig. 3(b). Temperature-dependent IPES measurements were also performed, showing that the spectral shape remained unchanged between low (15 K) and high (300 K) temperatures, as shown in Fig. 3(c). In the unoccupied state, the origin of the structure at the energy of $\omega_2$ and 5 - 8 eV is unclear because its energy position overlaps with the calculated Ce 4*f* and Nd 4*f* unoccupied states (see Fig. 4), and we therefore examine its possible origin in detail in the following section.

**Discussion**

To elucidate the origin of the electronic structures ($\omega_1$ and $\omega_2$) observed by ARPES and IPES in details, we performed DFT+DMFT calculations for $Nd_{1.857}Ce_{0.143}CuO_4$ (see also Methods for details). Following the exclusion of a high-energy-kink origin for the feature at $\omega_1$, consideration is given to alternative possibilities. Electron doping, introduced by substituting Nd with Ce, has been demonstrated to modify the electronic states in the $CuO_2$ plane. One potential explanation for the weak spectral weight at $\omega_1$ is that it may arise from Nd 4*f* and/or Ce 4*f* states. Nevertheless, the observation of the *f* electrons in proximity to $E_F$ would be challenging, given that the relaxation energies associated with *f*-electrons have been theoretically suggested to be of the order of several electron volts [45]. This substantial relaxation-energy value suggests that *f*-derived bands are expected to appear significantly away from $E_F$ in the spectroscopic measurements. Resonant photoemission studies have established that the Nd 4*f* (Ce 4*f*) exist at energies of 4–5 eV (3–4 eV) below $E_F$, far deeper than the hump structure observed at $\omega_1$ [46, 47], ruling out a Nd 4*f* and Ce 4*f* origin. Furthermore, even advanced theoretical calculations combining the local density approximation with dynamical mean-field theory and momentum-dependent self-energy (LDA+DMFT+$\Sigma_k$) fail to reproduce the ARPES spectral features near the $E_F$ in NCCO [48, 49]. In contrast, theoretical studies based on the extended *t*–*J*–*V* model have shown that the structures at $\omega_1$ in Fig. 2 naturally arise from coupling to optical plasmons [31, 33] rather than from localized 4*f* states. Within this plasmon-coupled spectrum, the energy band at $\omega_1$ appears as the incoherent plasmaron side band and plasmaron bands.

In the unoccupied states, the observed IPES spectra are well assigned by band structures derived principally from the Nd 4*f* and Ce 4*f* orbitals as shown in Fig. 4. The energy range of 5–8 eV is almost entirely governed by hybridized bands with a strong Nd 4*f* character as shown in Fig. 4. The theoretical predictions by *t*-*J*-*V* model indicate that coupling to charge fluctuations can give rise to the QP band structures, coherent side bands, in the range of 5–8 eV with a distinct dispersive feature in electron-doped cuprates [31]. In the present DFT+DMFT calculations, the energy of the predicted coherent side band overlaps with that of the localized bands originating from the Nd 4*f* orbitals. Consequently, to clarify the origin of the electronic structures observed at 5–8 eV in the IPES spectra in Fig. 3, it is essential to distinguish between the coherent side band and the Nd 4*f*-derived band. While the energy resolution of IPES is limited compared with that of ARPES, QP band structures with strong dispersion can still be detected. In practical, however, the peak position of the IPES spectra remains essentially non-dispersive in wide momentum space. Therefore, a strong peak structure would probably correspond to the Nd 4*f*-derived bands, and even if the coherent side band exists, it might not to be observed because its intensity is weaker than that of Nd 4*f*.

The origin of the peak or shoulder structure observed at approximately $\omega_2 \sim 2$–4 eV in the IPES experiment also shall be addressed. Our present DFT+DMFT calculations and incident-energy-dependent IPES measurements provided a unified understanding of the unoccupied electronic structure of NCCO. In the DFT+DMFT spectra, the Ce 4*f*-derived flat band robustly appears at approximately +1.5 eV, reflecting the strongly localized and nearly atomic character of the Ce 4*f* orbital. This sharp feature is consistent with the well-screened $4f^1$ excitation observed in $CeO_2$ and metallic α-Ce [50–52]. In contrast, Ce in NCCO acts primarily as an electron dopant and is known to be close to a $Ce^{4+}$ ($4f^0$) configuration [53]. For such a $4f^0$ ground state, the electron-addition $4f^1$ final state is expected to be shifted to substantially higher energies [45].

To confirm the presence and character of the Ce $4f^1$ spectral feature, we performed RIPES measurements, as shown in Fig. 5. RIPES is particularly powerful for detecting weak 4*f* signals owing to the small 4*f* cross section. Resonant photoemission spectroscopy enhances orbital-specific spectral weight at selected absorption edges, enabling a clear distinction between localized and hybridized electronic states through their characteristic resonance behaviors. The previous resonant ARPES combined with hybrid-functional DFT study has established that $CeO_2$ hosts a covalent O 2*p*–Ce 4*f* hybridized valence band, in sharp contrast to the nearly nondispersive and localized 4*f* state in *c*-$Ce_2O_3$ [54]. These contrasting behaviors highlight that RIPES is suited to determine whether the Ce 4*f* state in NCCO remains an impurity-like, localized $4f^0 \rightarrow 4f^1$ final state or participates in hybridization to form more itinerant unoccupied bands.

As demonstrated in Fig. 5(a), when the incident energy was tuned to the Ce 4*d*→4*f* resonance at $E_i$ = 120 eV in NCCO, a sharp and strongly enhanced peak appeared at +2.5 eV. Furthermore, the RIPES spectrum taken at $E_i$ = 120 eV shows an asymmetric shape as shown by a triangle. This finding suggests that the resonance behavior unambiguously identifies the 2.5-eV feature as the coherent Ce $4f^1$ final state, and that another electronic structure also exists in close proximity to the Ce $4f^1$ state. This is due to the fact that the Ce $4f^1$ final state are expected to show a symmetric spectral shape. The 0.5-eV shift relative to the calculated 1.5-2 eV initial state is naturally explained by the expected relaxation energy of the Ce $4f^1$ configuration [45]. As shown in Fig. 5(b), the RIPES spectrum was fitted with several Lorentzian functions, from which we extracted the QP-band component represented by the cyan-colored Lorentzian at the energy of 3.5 eV. We attribute this feature to an emergent QP band structure, which exhibits only a weak enhancement at the Ce 4*d*–4*f* resonance, in sharp contrast to the strongly resonant Ce $4f^1$ final state. In general, an emergent QP band arising from quantum charge fluctuations is not expected to exhibit any resonance. However, the weak resonant enhancement of the electronic structure at 3.5 eV may indicate that the $CuO_2$ -derived charge fluctuation is involved in the screening of the localized Ce $4f^1$ final state. Note that this should be regarded as dynamical screening, rather than ground-state hybridization between the Ce 4*f* states and the conduction bands. Taken together, the theoretical and RIPES evidence indicates that the electronic structure at 3.5 eV in NCCO is more consistently explained as a plasmaron-related excitation and the Ce $4f^1$ final state. This interpretation resolves the apparent discrepancy between the Ce 4*f* level at 1.5 - 2 eV by DFT+DMFT calculations and the experimental IPES spectrum and highlights the essential role of long-range charge dynamics and the relaxation energy in shaping the unoccupied electronic structure of electron-doped cuprates. The present study contributes to the development of a theoretical framework that clarifies the role of charge fluctuations in both electron- and hole-doped cuprate superconductors. We note that a recent ARPES study on transition-metal oxides $SrVO_3$ have demonstrated that dynamical screening and plasmonic coupling can reshape the electronic structure in the occupied state as a weak replica band and modifying QP bands [40]. This experimental evidence highlights that electron-plasmon coupling is a general and essential ingredient for an understanding of correlated oxides. The electronic structures observed by ARPES and IPES on NCCO in the present study can be interpreted within this broader framework, where collective charge fluctuations in the $CuO_2$ planes contribute to the dynamical screening of both occupied and unoccupied states.

Although the present ARPES and IPES measurements on optimally doped NCCO already provide substantial insight into the emergent QP band structures by charge fluctuations, one expects that systematic studies across different doping levels may reveal these features even more clearly. This may in turn allow their identification in ARPES and IPES measurements without interference from the Cu 3*d*, O 2*p*, Nd/Ce 4*f* bands. Because the energies of both the incoherent plasmaron and the coherent side bands are expected to scale with the optical plasmon energy, which changes with carrier concentration in both electron- and hole-doped cuprates [23, 55, 56]. Consequently, doping-dependent ARPES and IPES experiments may provide further definitive evidence regarding the nature of the charge fluctuation-derived QP band structure.

**Conclusions**

Our combined ARPES, IPES, and DFT+DMFT studies reveal not only the electronic structures below and above Fermi level but also those induced by coupling to charge fluctuations in the electron-doped cuprate superconductor. In the occupied states, the ARPES spectra reveal the electronic structure around –1 eV, which may correspond to incoherent plasmarons. In the unoccupied states, the DFT+DMFT calculations support that the electronic structure of 5 – 8 eV probably originates from the Nd 4*f* states. Furthermore, we experimentally identified a clear resonant enhancement of the Ce 4*f* states at 2.5 eV, which arises from the Ce $4f^1$ final state, revealing a discrepancy from the calculated Ce 4*f* level. In addition, the RIPES spectrum exhibits a shoulder structure, and we found that an incoherent plasmaron-related QP band structure coexists in close proximity to the Ce $4f^1$ final-state.

## Methods

### Fabrication of samples and experimental conditions

High-quality single crystals of optimally doped NCCO ($x$ = 0.15, $T_c$ ~ 25 K) were synthesized using the traveling floating-zone method and the $T_c$ was determined by measuring the magnetic susceptibility. The samples were clamped to the sample holder with silver paste and cleaved *in-situ* for the ARPES and IPES experimental chambers at a pressure of ~$5 \times 10^{-9}$ Pa.

### Angle-resolved photoemission spectroscopy

ARPES experiments were performed at BL7U ($h\nu$ = 16.7 eV) of UVSOR-III Synchrotron, Institute for Molecular Science. The temperature and energy resolution of the ARPES measurements was at $T$ = 7 K and 8–10 meV, respectively. The measurements were conducted utilizing $s$-polarized light and the MBS-A1 electron analyzer. ARPES data are processed by Igor Pro 6.35 software.

### Inverse photoemission spectroscopy

IPES ($E_i$ = 50 - 125 eV) experiments were performed at Institute for Synchrotron Radiation Science, Hiroshima University (HiSOR). The temperature and energy resolution of the IPES measurements were at 15 K - 300 K and ~ 300 meV, respectively. The angle of directions between the incidence electrons and emitted photons was fixed to be 45° and the AR-IPES spectra along (0, 0) - (π, π), (0, 0) - (π, 0), and (π, 0) - (π, π) directions were measured with changing the incidence angle of electrons relative to the sample surface nominal in steps of 1° because the detector is a single channel. The temperature of samples was controlled by using the closed-cycle He refrigerator and heater to measure the spectra at 300 K. The angular divergence of the electron beams to be at most 4°, corresponding to the momentum resolution of 0.24 Å$^{-1}$. An energy position of $E_F$ of the IPES spectra was determined from the Au spectra. IfPES data are processed by Igor Pro 6.35 software.

### DFT calculations

Calculations of the electronic structure were performed using OpenMX version 4.0, which employs norm-conserving pseudopotentials and localized numerical pseudo-atomic orbitals [60, 61]. We used an ordered 49-atom explicit-substitution supercell with the composition $Nd_{13}Ce_1Cu_7O_{28}$, corresponding to $x$ = 1/7 = 0.142857, which is close to the Ce concentration ($x$ ~

0.15) examined in this study. The lattice and atomic coordinates were kept fixed, and no ionic relaxation was performed. Exchange and correlation were treated within the Perdew-Burke-Ernzerhof generalized-gradient approximation [57]. Spin-orbit coupling was included in a non-collinear calculation using fully relativistic pseudopotentials. The numerical pseudo-atomic-orbital basis sets were Nd8.0-s3p2d2f1, Ce8.0-s3p2d2f1, Cu6.0S-s3p2d1, and O6.0-s2p2d1. A real-space integration cutoff of 220 Ry and a 4 × 4 × 4 k-point mesh were used.

**DMFT calculations**

Local electronic correlations were treated within a non-charge-self-consistent DFT+DMFT scheme [58, 59] using the Hubbard-I approximation [65]. The impurity problems were solved using the TRIQS framework [62]. The correlated subspace was constructed from the Nd and Ce 4*f* pseudo-atomic orbitals using a global symmetric Löwdin orthonormalization [63]. Each rare-earth site was represented by a 14-dimensional spin-orbital subspace. The one Ce site and all thirteen Nd sites were treated separately, without a representative-site reduction. Correlations were applied only to the Nd and Ce 4*f* shells. A density-density Kanamori interaction with $U = 9.7$ eV and $J = 0.5$ eV was used for both the Nd and Ce 4*f* shells [66], and the inverse temperature was $\beta = 40$ eV$^{-1}$.

**Treatments of double counting**

Shell-dependent fixed double-counting potentials were used rather than a self-consistently updated occupancy-based correction. The reference potentials were evaluated using the Held average interaction appropriate for the density-density Kanamori form [64]. For the Ce 4*f* shell, $N_{\mathrm{ref}} = 0.16$ and an additional double-counting shift of $\Delta_{\mathrm{DC}} = 0$ were used. Here, $N_{\mathrm{ref}}$ is a fixed double-counting reference and is not an enforced impurity occupancy. The value $N_{\mathrm{ref}} = 0.16$ was motivated by the reported effective Ce valence of approximately +3.84 in NCCO [67]. For the Nd 4*f* shells, the nominal trivalent reference $N_{\mathrm{ref}} = 3$ was used together with an additional double-counting shift of $\Delta_{\mathrm{DC}} = -1.6$ eV. This shift was introduced as a spectroscopically guided alignment of the localized Nd 4*f* energy window. Consequently, the absolute positions of the calculated Nd 4*f* structures should be regarded as semi-quantitative rather than as parameter-free first-principles predictions. The calculation is used primarily to assist the orbital assignment of rare-earth-derived states.

**Data availability**

The data that support the findings of this study are available from the corresponding author upon request.

**Code availability**

The computer code that shows the calculations of this study is available from the corresponding author upon request.

**Funding statement**

S. I. acknowledges the support from the HIRAKU-Global Program, which is funded by MEXT's "Strategic Professional Development Program for Young Researchers." D. S. thanks to funding from the Max Planck–UBC–UTokyo Centre for Quantum Materials and the Canada First Research Excellence Fund (CFREF), Quantum Materials and Future Technologies. This work was supported by the Japan Society for the Promotion of Science (JP24K06961).

**Acknowledgements**

The authors thank H. Yamase, T. Yoshida, and D. Otsuki for their valuable discussions. The ARPES results were obtained at the UVSOR-III Synchrotron (Proposal No. 23IMS6854) and the IPES experiments were performed at Research Institute for Synchrotron Radiation Science (Proposal Nos. 25AG006 and 24AU022). We also acknowledge the OpenMX development team and the developers of TRIQS and triqs_hubbardI for providing and maintaining the software used in this work.

**Author contributions**

S. I. and K. S. conceived and coordinated the research. D.S. grew the high-quality single crystals of NCCO. ARPES experiments were performed by Y. T., Y. M., Y.K., and S.I. and supported by K. T. IPES experiments were performed by H. Y., H. S., and S.I., and supported by M. A. In the theoretical part of this study, H. Y. performed the DFT and DMFT calculations. H. Y., S. K., and S. I. wrote the manuscript with input from all the authors.

**Competing interests**

The authors declare no competing interests.

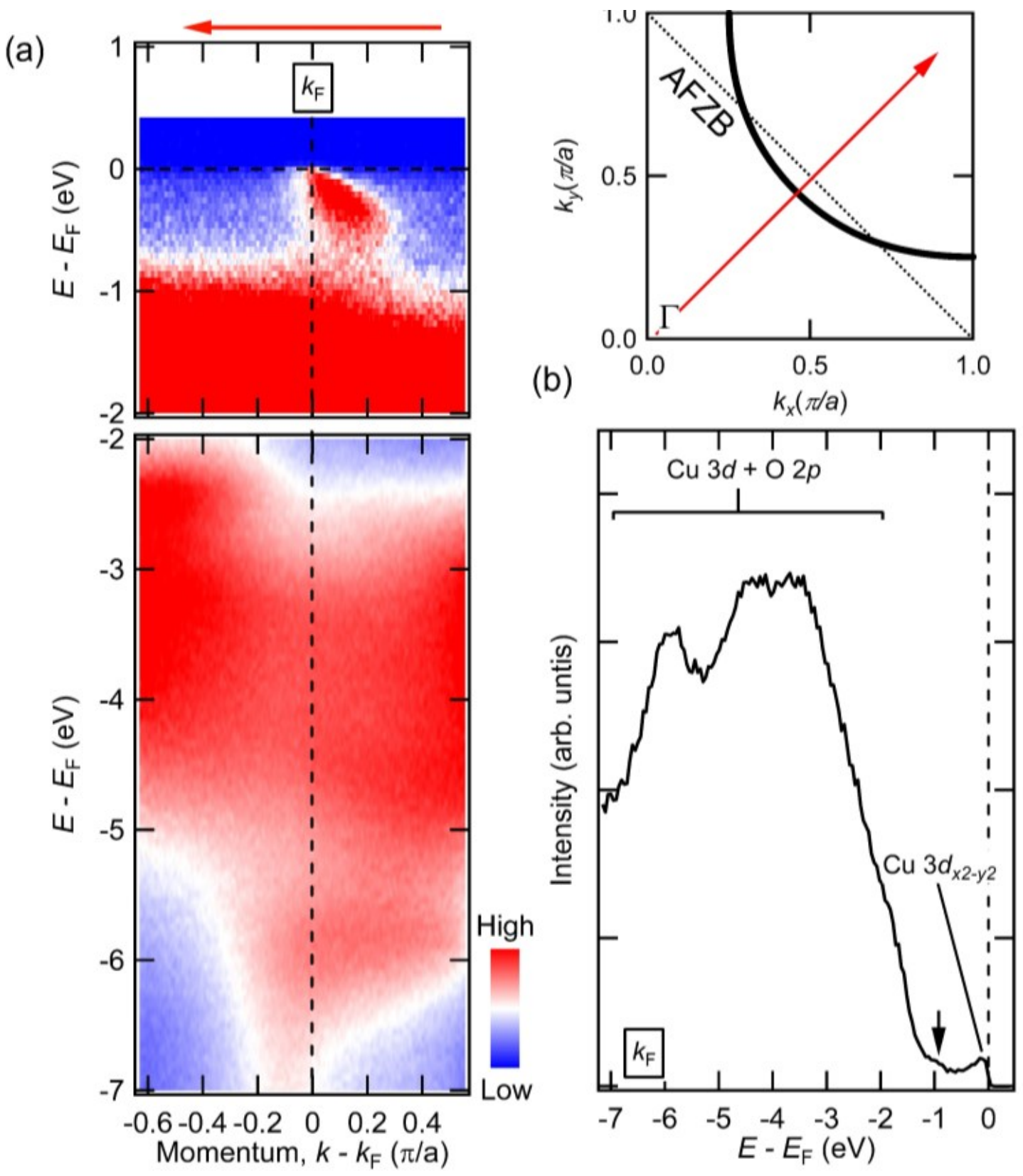

**Fig. 1 Energy–momentum (*E*–*k*) intensity plots for NCCO.** (a) ARPES intensity plot as a function of energy and momentum. The red arrow indicates the momentum direction as shown in the schematic Fermi surface (black solid curve). To highlight the weak spectral weight near the Fermi level ($E_F$) associated with the nodal QP band, the ARPES intensity plots for deep binding energies are shown in the separate panel. AFZB indicates the antiferromagnetic zone boundary. (b) Energy distribution curve (EDC) at the Fermi momentum ($k_F$), corresponding to the vertical and black dotted line in panel (a).

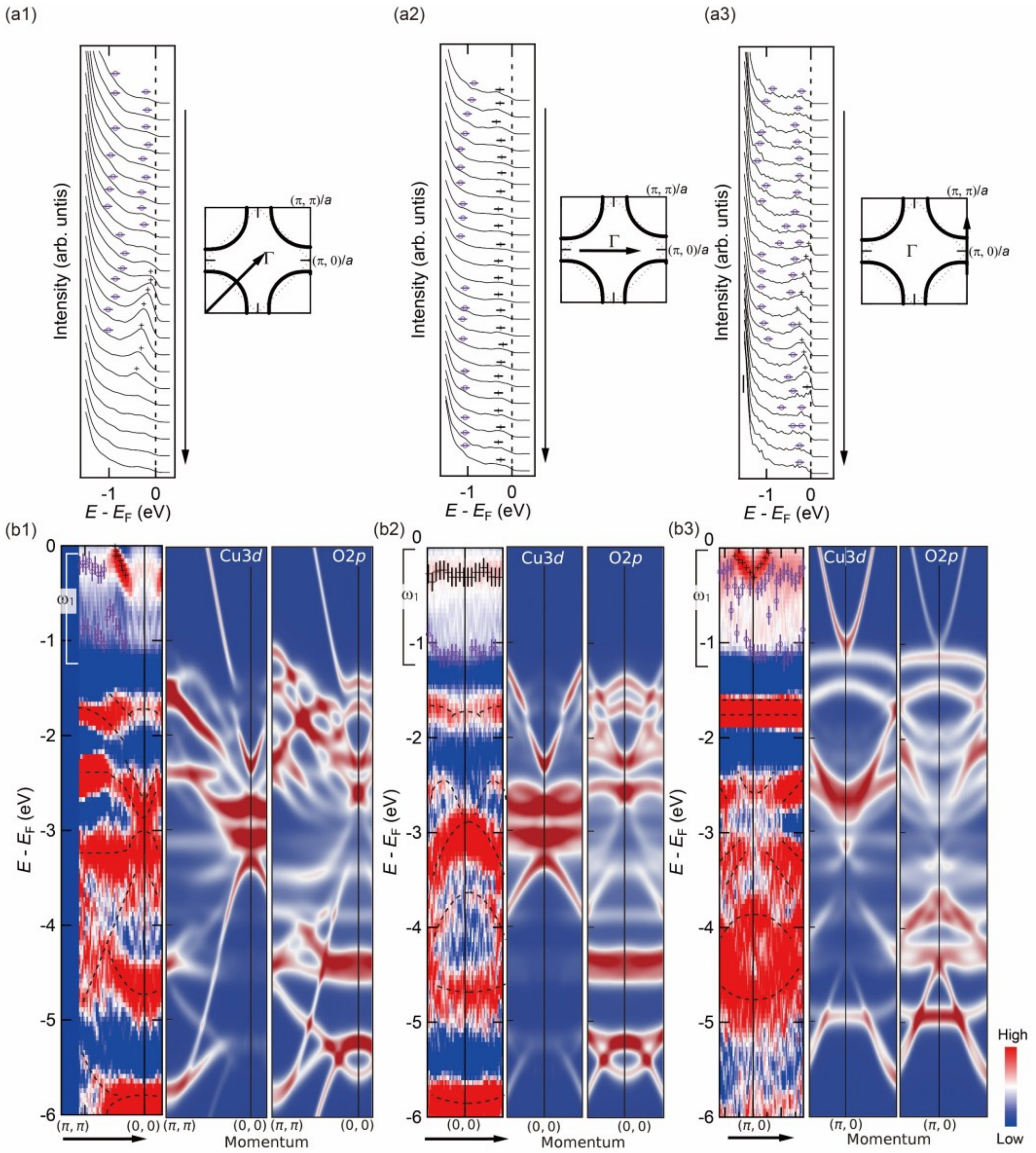

**Fig. 2 ARPES spectra along high-symmetry momentum directions and band structure calculations using DFT+DMFT for $Nd_{1.857}Ce_{0.143}CuO_4$ in the occupied states.** Panels (a1), (a2), and (a3) are the energy-distribution curves (EDCs) along (0, 0) – (π, π), (0, 0) – (π, 0), and (π, 0) – (π, π), respectively. Black crosses in panels (a1), (a2), and (a3) indicate peak positions and/or shoulder structures, corresponding to the Cu 3*d* and O 2*p* bands. The electronic structure observed in this measurement is indicated by the purple-circled markers, and it extends broadly from the $E_F$ down to approximately –1 eV ($\omega_1$). (b1) - (b3) The second-derivative *E*-*k* plots and calculated ARPES images corresponding to the experiments. The electronic structure of Cu 3*d* and O 2*p* is strongly observed as shown by a black dotted curve in the panels (b1) – (b3). The spectral intensity of the second-derivative analysis of the *E*-*k* plots is inverted for the purpose of comparison with the calculation. The QP bands simulated by DFT and DMFT for the Cu3*d* and O 2*p* states are also shown in the corresponding momentum. The momentum directions for the *E*-*k* plots are indicated by a black straight arrow on the schematic Fermi surface and each panel. The error bar is an uncertainty of the peak and shoulder positions of the ARPES spectra.

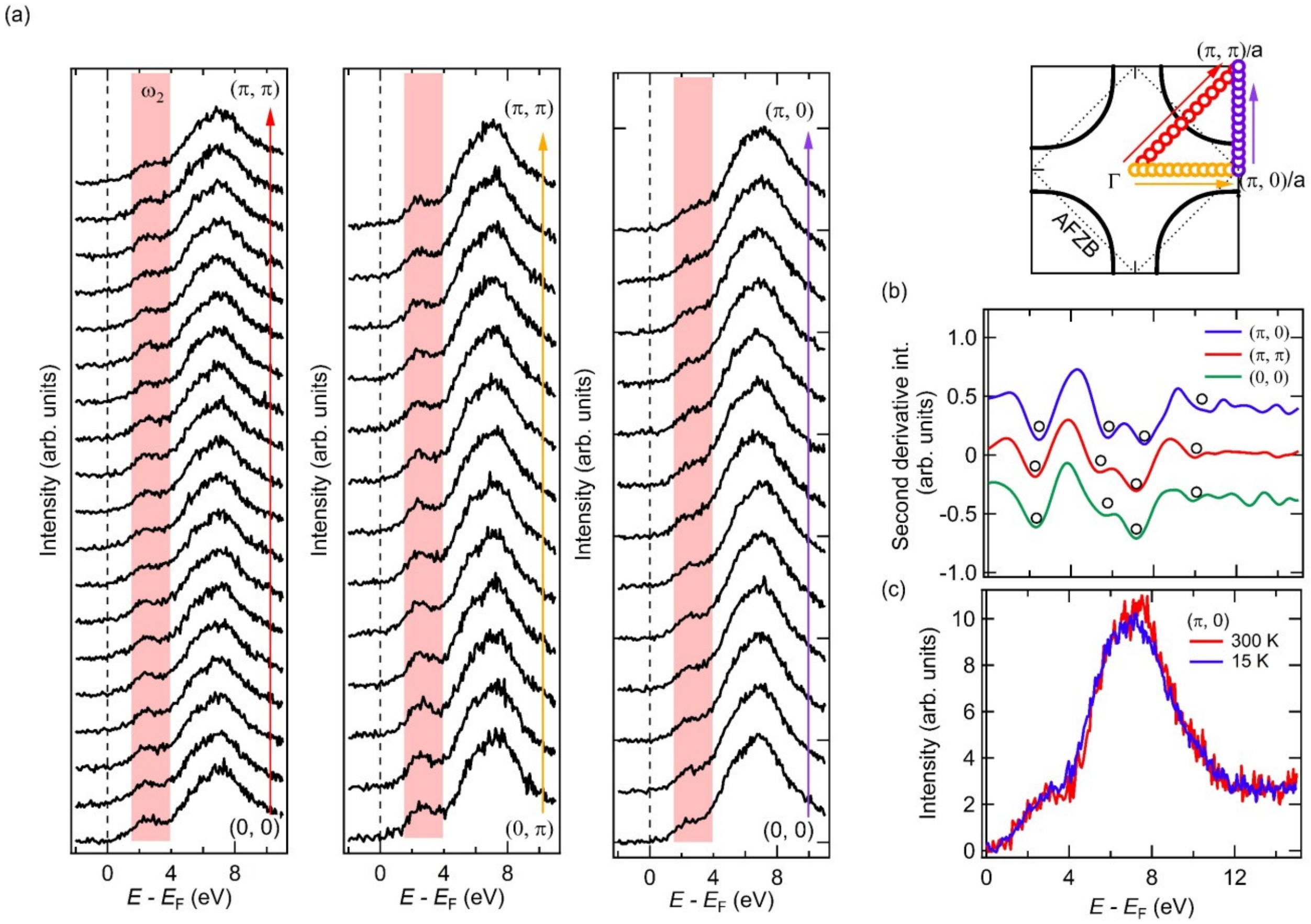

**Fig. 3 Angle-resolved inverse photoemission spectroscopy (AR-IPES) spectra of NCCO.** (a) AR-IPES spectra taken at $E_i$ = 50 eV along the momentum directions (0, 0) – (π, π), (0, π) – (π, π), and (0, 0) – (π, 0) and momentum positions are indicated by arrows and markers on the schematic Fermi surface, respectively. (b) Second-derivative AR-IPES spectra with respect to energy. Peak (inflection) positions are marked with circles. (c) Temperature dependence of IPES spectra taken at $T$ = 15 K and 300 K.

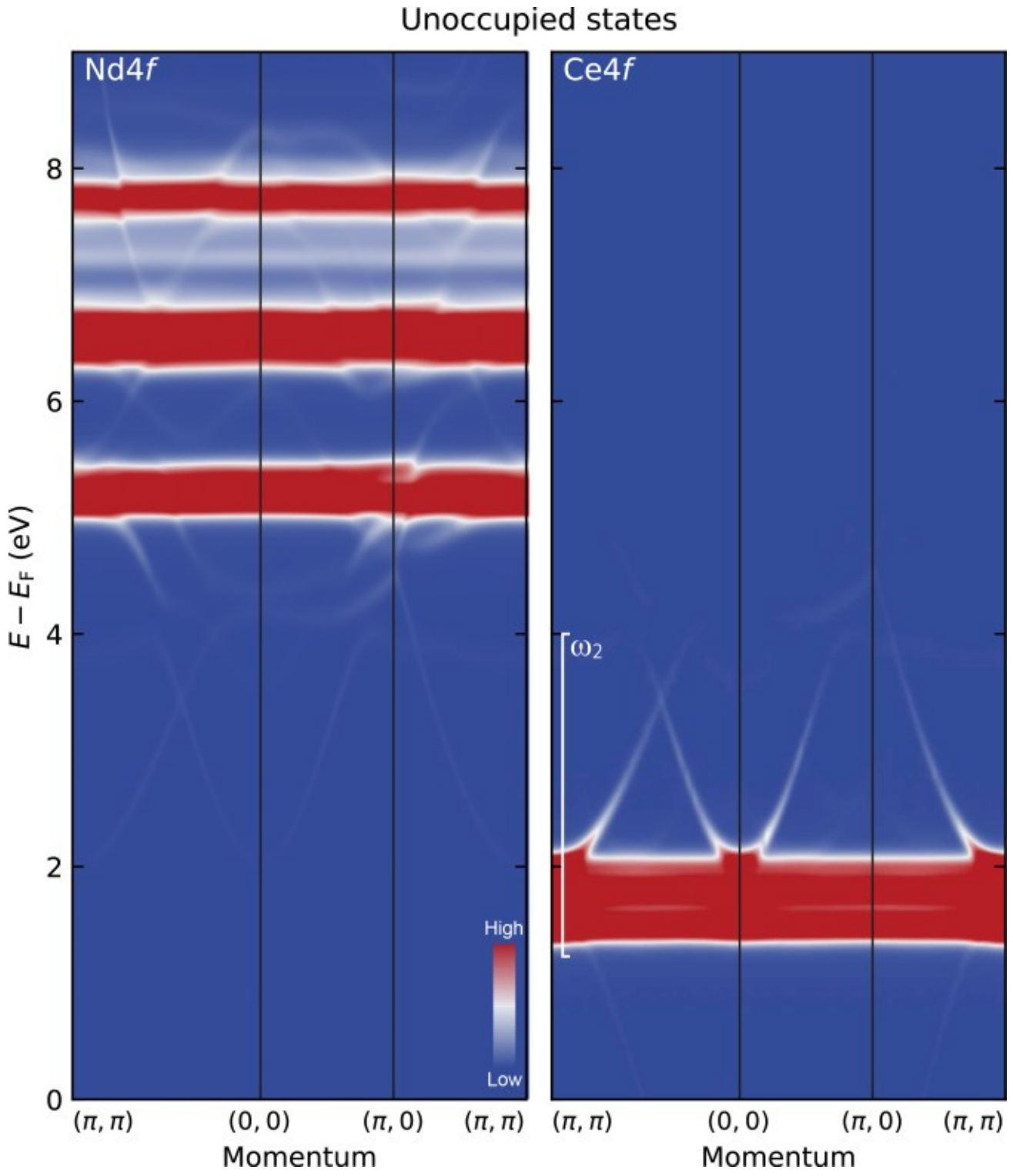

**Fig. 4 Band structure calculations using DFT+DMFT for $Nd_{1.857}Ce_{0.143}CuO_4$ in the unoccupied states.** A strong intensity from Nd 4*f* orbitals dominate the energy range of 5 − 8 eV, while the intensity corresponding to the density of states of Ce 4*f* is shown around 1.5-2 eV. See also Methods for details of the calculations.

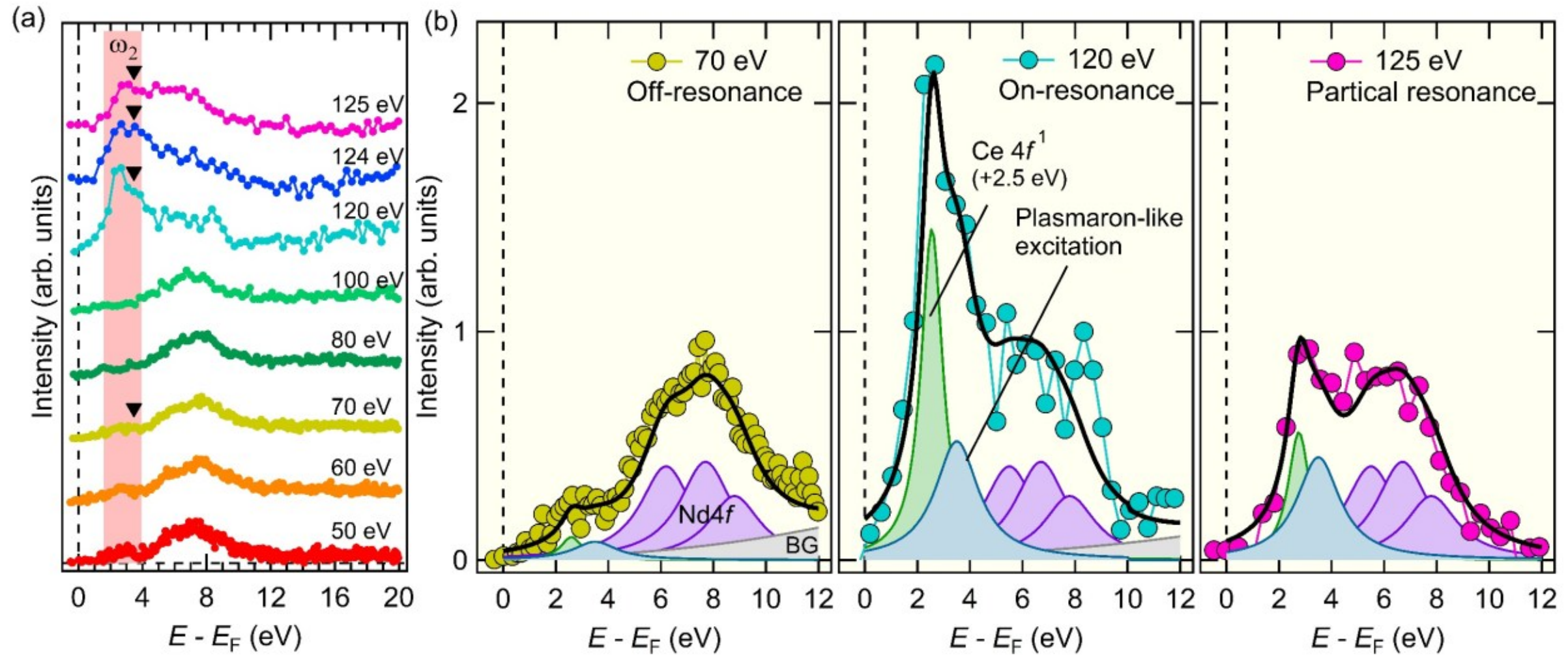

**Fig. 5 IPES spectra of NCCO across the Ce 4*d*-4*f* resonance at Γ.** (a) IPES spectra measured at selected incident electron energies ($E_i$ = 50-125 eV). Red area shows the energy range of $\omega_2$. The energy of a shoulder structure on the spectrum is shown by triangle. (b) The broad 2-4 eV structure as labeled as the plasmaron-like excitation is weak at data of $E_i$ = 70 eV (off-resonance). At $E_i$ = 120 eV, the sharp peak at 2.5 eV is strongly observed (on-resonance) and is assigned to the Ce $4f^1$ final state. In addition, the RIPES spectrum of the Ce $4f^1$ state has the shoulder structure at the data of $E_i$ = 120 eV. The IPES spectra are fitted using several Lorentzian functions, and the Nd 4*f* orbitals are fitted using the minimum number of Lorentzian functions (purple) necessary to roughly reproduce the electronic structure indicated by the present DFT+DMFT calculations. Nearly identical parameters are used for each Lorentzian fitting of the Nd 4*f* orbitals, and the same parameters are also used for the peak positions and widths to reproduce the structures of the Ce 4*f* and plasmaron-like excitations for each fitting. The black curve represents the spectrum obtained by combining the Lorentzian function used in the fit with the background (BG).